\documentclass[numbers=endperiod]{scrartcl}
\usepackage{ebgaramond}

\usepackage{graphicx}

\usepackage{float}
\usepackage{amsmath}
\usepackage{bbm}
\usepackage{caption}
\usepackage{multirow}
\usepackage{breakcites}
\usepackage{enumitem}
\usepackage{authblk}
\usepackage[symbol]{footmisc}

\usepackage{titlesec}

\titleformat{\section}
  {\normalfont\scshape\Large\bfseries}
  {\thesection}{1em}{}
\titleformat{\subsection}
  {\normalfont\scshape\large\bfseries}
  {\thesubsection}{1em}{}
 \titleformat{\subsubsection}
  {\normalfont\scshape\bfseries}
  {\thesubsubsection}{1em}{}
 \titleformat{\paragraph}
{\normalfont\scshape\normalsize\bfseries}
{\theparagraph}{1em}{}
 \titleformat{\subparagraph}
{\normalfont\scshape\normalsize\bfseries}
{\thesubparagraph}{1em}{}

\setkomafont{title}{\normalfont\huge\scshape}
\setkomafont{subtitle}{\normalfont\LARGE\scshape}

\usepackage[english]{babel}

\title{\textsc{Tit for Tattling: Cooperation, communication, and how each could stabilize the other}}
\author[1]{Victor Vikram Odouard\footnote[1]{Author to whom correspondence should be addressed: Victor Vikram Odouard, Santa Fe Institute (Cowan Campus) 1399 Hyde Park Road, Santa Fe, New Mexico 87501, U.S.A. E-mail: vo47@cornell.edu.}}
\author[1]{Michael Holton Price}

\affil[1]{\begin{normalsize}Santa Fe Institute, 1399 Hyde Park Road, Santa Fe, NM 87501\end{normalsize}}

\begin{document}

\begin{center}
    \makeatletter
    \begin{huge}
        \@title \\[0.2em]
    \end{huge}
    \begin{Large}
        \emph{by}\\
      \@author \\[2em]
    \end{Large}
\end{center}

%TC:ignore
\begin{center}
    \textsc{Abstract}
\end{center}
Indirect reciprocity is a mechanism by which individuals cooperate with those who have cooperated with others. This creates a regime in which repeated interactions are not necessary to incent cooperation (as would be required for direct reciprocity). However, indirect reciprocity creates a new problem: how do agents \emph{know} who has cooperated with others? To know this, agents would need to access some form of \emph{reputation} information. Perhaps there is a communication system to disseminate reputation information, but how does it remain truthful and informative? Most papers assume the existence of a truthful, forthcoming, and informative communication system; in this paper, we seek to explain how such a communication system could remain evolutionarily stable in the absence of exogenous pressures. Specifically, we present three conditions that together maintain both the truthfulness of the communication system and the prevalence of cooperation: individuals (1) use a norm that rewards the behaviors that it prescribes (an aligned norm), (2) can signal \emph{not only} about the actions of other agents, but also about their truthfulness (by acting as third party observers to an interaction), and (3) make occasional mistakes, demonstrating how error can create stability by introducing diversity.
% RESOLVED: we should probably mention metasignaling and observation in the abstract. We should have space for this as I removed some of the background material.

\begin{center}
\end{center}

% I think the highlights read better being capitalized, but feel free to reverse this choice.
\begin{center}
    \textsc{Highlights}
\end{center}
\begin{itemize}
    \item We find a state in which cooperation and communication stabilize each other
    \item A signaling system intended initially for stabilizing cooperation can be co-opted to also stabilize \emph{itself}
        % TODO: consider rewording the preceding bullet (since in the full model any intent of the previous model is irrelevant)
    \item Errors in cooperation decisions bring out unexpressed traits that would otherwise allow for the invasion and demise of cooperative equilibria
    \item Norms, when followed by gossipers, behave differently from norms followed by direct observers. In the latter case, the norms of invaders affect only their own actions, but in the former, they can affect the actions of conforming agents with whom they gossip
\end{itemize}

%TC:endignore
\section{Introduction}
%Explanations for altruism, such as kin selection, reciprocity, indirect reciprocity, punishment, and genetic and cultural group selection, typically involve mechanisms that \textit{make altruists more likely to benefit from the altruism of others}. In the case of kin altruism and reciprocity, individuals use private information to identify targets of their altruism. In the case of indirect reciprocity, where individuals cooperate with high-reputation individuals, outside information is required: unless agents can directly observe every interaction, communication between agents disseminates reputation information. But most accounts of indirect reciprocity take as given a truthful and misunderstanding-free communication system.

Explanations for the evolution of altruism (helping others at a personal cost) including kin altruism \cite{hamilton-the-genetical-evolution-of-social-behavior-1964}, reciprocity \cite{trivers-the-evolution-of-reciprocal-altruism-1971}, indirect reciprocity \cite{nowak-evolution-of-indirect-reciprocity-by-image-scoring-1998}, punishment \cite{boyd-the-evolution-of-altruistic-punishment-2003}, and group selection \cite{bowles-the-coevolution-of-individual-behaviors-and-social-institutions-2003}, all ensure that altruists receive a second-order benefit that compensates for the initial cost: for instance, they may be more likely (compared to non-altruists) to benefit from altruism, or perhaps, less likely to suffer from punishment \cite{henrich-cultural-group-selection-coevolutionary-processes-and-large-scale-cooperation-2004, queller-quantitative-genetics-inclusive-fitness-and-group-selection-1992}).

Under kin altruism, if an agent is altruistic, their kin are more likely to be altruistic, producing a second-order benefit by making them more likely to benefit from the altruism of others \cite{queller-a-general-model-for-kin-selection-1992}. Direct reciprocity means altruists are more likely to have their favors returned \cite{brown-the-evolution-of-social-behavior-by-reciprocation-1982}. Under indirect reciprocity, individuals have their favors returned even by third parties, without the need for repeated interactions \cite{panchanathan-a-tale-of-two-defectors-the-importance-of-standing-for-the-evolution-of-indirect-reciprocity-2003}. Altruistic punishers provide a second-order \emph{relative} benefit to altruists by punishing non-altruists \cite{boyd-the-evolution-of-altruistic-punishment-2003} (or, possibly, a second-order benefit to themselves by coercing compatriots into altruistic behavior \cite{boyd-punishment-allows-the-evolution-of-cooperation-or-anything-else-in-sizable-groups-1992}). And with group selection, groups that have high numbers of altruists are the ones most likely to spread, so most altruists will be concentrated in groups with high densities of altruists \cite{odouard-polarize-catalyze-stabilize-conscience-and-the-evolution-of-cooperation-2021}. In all these ways, altruism produces a second-order benefit compensating for its first-order cost, making it potentially evolutionarily viable.

However, large groups hinder the ability of individuals to direct altruism towards altruists, because (1) their interactions are more likely to occur between non-kin, weakening kin selection, (2) they cluster closer to the population average, reducing between-group variance that group selection relies on \cite{smith-group-selection-1976, nowak-five-rules-for-the-evolution-of-cooperation-2006}, and (3) they make repeated interactions rare, weakening reciprocity \cite{brown-the-evolution-of-social-behavior-by-reciprocation-1982, henrich-cultural-group-selection-coevolutionary-processes-and-large-scale-cooperation-2004}. 

Indirect reciprocity mitigates the large-group problem, as individuals do not need to interact directly to know each other's reputations. But vitally, if direct observation of every interaction is infeasible, communication is required to disseminate reputation. 

Most models simply assume the existence of an informative and truthful communication system. In this paper, we do not simply make this assumption; instead, we consider the conditions under which communication and cooperation can remain stable. Specifically, we ask 

\begin{quote}
    \emph{Under what conditions does the interaction between signaling and cooperation stabilize both high levels of (altruistic) cooperation and truthful, informative, and forthcoming (i.e. effective) communication?}
\end{quote}

% TODO: the following sentence (and subsequent material) is the center of our paper. Perhaps there's a way to better highlight this.
\noindent In this paper, we answer this question by finding a stable, highly cooperative, and effectively communicative state (definitions in Section~\ref{section:model1-definitions}) in an indirect reciprocity model, with discounting, (Section~\ref{section:model1-the-model}) where agents evolve both rules for how to
\begin{enumerate}[label=\roman*.]
\item \emph{act} in a prisoner's dilemma (i.e., behavior, Section~\ref{section:model1-strategies-action}),  \emph{and} 
\item \emph{communicate} about the actions of others (i.e., language, Section~\ref{section:model1-strategies-signal}). 
\end{enumerate}
Under the conditions that allow for a stable cooperative-communicative state, agents
%In such a model, `lying' is expressed by individuals with communication strategies that deviate from the norm.
\begin{enumerate}
    \item \textit{act and signal according to an `aligned' norm}, that is, a norm that rewards the actions that it prescribes. That is, if the norm prescribes defection against agents in `bad moral standing', it must also reward defection against those agents (Section~\ref{section:model1-strategy-focal-strategy}).
    \item \textit{occasionally deviate from their strategy}, which prescribes particular actions in particular situations. If everyone employs the same strategy, the environment becomes homogenized to the point where agents become unprepared for novel threats. Deviations keep the environment sufficiently variable that agents can remain prepared (Section~\ref{section:model2-strategies-and-error}). 
    \item \textit{exert normative pressure on each other's signals}, which allows benefits not only to be disproportionately distributed to cooperators in the prisoner's dilemma, but \emph{also} to truthful communicators (Section~\ref{section:model2-strat-meta-signaling}).

\end{enumerate}

% TODO: circle back to the following summary to update it with final sections, etc.

In Section~\ref{section:background}, we will provide background on relevant concepts, which some may choose to skip. In section Section~\ref{section:model1}, we present a basic model of indirect reciprocity, showing why communication does not remain stable in such a regime. We then present an altered model in Section~\ref{section:model2} that \emph{does} have a stable, cooperative state. Finally, in Section~\ref{section:discussion}, we discuss the broader significance of our results.

\section{Background}\label{section:background}
\subsection{Indirect reciprocity stabilizes cooperation in large groups...} \label{section:background-indirect-reciprocity}
Indirect reciprocity can help to explain pairwise altruism when acts are unlikely to be directly reciprocated, such as when groups are large, because \emph{third parties} can withhold cooperation specifically from those with bad reputations (creating targeted punishment \cite{boyd-the-evolution-of-reciprocity-in-sizable-groups-1988}). It has been observed experimentally, where people often cooperate disproportionately with those who have cooperated with others when there is no possibility of having their act directly reciprocated, suggesting that reputation is not just an aid for estimating one's own future payoff \cite{wedekind-cooperation-through-image-scoring-in-humans-2000}. And ethnographically, some societies tolerate stealing from families with bad reputations \cite{henrich-the-origins-and-psychology-of-cooperation-2021, bhui-how-exploitation-launched-human-cooperation-2019}.

A strategy is a set of rules for acting and signaling. Sometimes we refer to a strategy as a `norm'; we do this when the strategy  encodes a set of rules that is followed by the large majority of a social group. The simplest norm in indirect reciprocity is image-scoring, which says that `cooperate with cooperators, defect with defectors'  \cite{nowak-evolution-of-indirect-reciprocity-by-image-scoring-1998}. However, this norm punishes those who defect against defectors, even though this is precisely (1) what the norm prescribes and (2) what makes defection costly \cite{nowak-evolution-of-indirect-reciprocity-2005}.

A more nuanced approach might be to introduce the notion of \textit{standing} (defined recursively: someone is in good standing if they cooperated, or if they defected against someone in bad standing), and to cooperate with agents if and only if (iff) they are in good standing \cite{ohtsuki-the-leading-eight-social-norms-that-can-maintain-cooperation-by-indirect-reciprocity-2005}. \textit{Stern judging}, similar to standing, additionally puts those who cooperate with those in bad standing into bad standing  \cite{pacheco-stern-judging-a-simple-successful-norm-that-promotes-cooperation-under-indirect-reciprocity-2006}. 

Both standing and stern judging create an incentive to heed an opponent's reputation when choosing how to act, unlike with image scoring case, where cooperation is good and defection is bad, regardless of your opponent's reputation \cite{leimar-evolution-of-cooperation-through-indirect-reciprocity-2000, panchanathan-a-tale-of-two-defectors-the-importance-of-standing-for-the-evolution-of-indirect-reciprocity-2003}. In part for this reason, these strategies have been shown capable of stabilizing states of high cooperation -- but all under the assumption that agents have perfect information about how others act, either by direct observation, or a completely truthful communication system.

\subsection{...but relies on effective communication}
The stability of honest communication systems usually rely on some kind of pressure on the signaler to be truthful \cite{grafen_1990, oliphant-the-dilemma-of-saussurean-communication-1996}: for example, when there is a common interest between receiver and signaler \cite{blume-evolution-of-communication-with-partial-common-interest-2001}, when signals are costly \cite{gintis-costly-signaling-and-cooperation-2001}, when there is a cost \text{differential} between truthful and non-truthful signaling (even if the equilibrium is cost-free)\cite{lachmann-cost-and-conflict-in-animal-signals-and-human-language-2001}, when the interaction is a coordination game (where both parties benefit from knowing the world-state of the other) \cite{mcelreath-shared-norms-and-evolution-of-ethnic-markers-2003, young-social-norms-and-economic-welfare-1998}, or when direct observation and partner-choice supplement the signals themselves \cite{robinson-arnull-moral-talk-and-indirect-reciprocity-direct-observation-enables-the-evolution-of-moral-signals-2018}. 

In its basic form, communicating the reputations of others places no such pressure on the signaler. Solutions have been offered: for instance, a truth-for-truth reciprocity system can be successful in sparking the rise of a truthful communication system \cite{oliphant-the-dilemma-of-saussurean-communication-1996}. The problem here, however, is that reputations and indirect reciprocity become important precisely when repeated interactions, on which reciprocity relies, become unlikely. But could an \emph{indirect} reciprocity mechanism enforce truthfulness (Section~\ref{section:model2-strat-meta-signaling})?

Truthfulness is not the only requirement of the communication system -- agents also need to be forthcoming, that is, actually share the information they have, despite, for instance, fear of reprisal. While in this model there is no mechanism directly representing such reprisal\footnote{For more on this, see \cite{wiessner_2009}, who has examined how well the assumptions of games-theoretic formulations like the ultimatum and dictator games align with living social norms. We might expect similar alignments and deviations for our communication model.}, it is still important to mitigate against it as an exogenous risk by making signal-withholding costly. The stable strategy deals with this by treating failures to signal in the same way it treats lying (Section~\ref{section:model2-strat-meta-signaling}).  

\section{Model 1: a first pass}\label{section:model1}
In this section, we will define Model 1, which will provide us insight into what conditions are required for a stable state of altruistic cooperation and effective communication. It provides a necessary foil for interpreting our primary model (Model 2; Section~\ref{section:model2}). Classic models of indirect reciprocity lock the signaling system, restricting an agent's strategies to the \emph{actions} they take (whether to cooperate or not). Our first model will remedy this by allowing for agency in both \emph{action} and \emph{signaling}.

\subsection{Definitions} \label{section:model1-definitions}
\subsubsection{Effective communication} \label{section:effective-communication}
We say that a communication system is a set of mappings from meanings to symbols, and we define an \textit{effective} communication system as one that is

\begin{enumerate}
    \item \textit{uniform}, that is, everyone abides by the same mapping from meanings to symbols,
    \item \emph{forthcoming}, that is, agents never fail to signal when they possess relevant information, and
    \item \textit{informative}, that is, everyone's mapping distinguishes between at least two meanings (in our case, the `meanings' are facts about the actions of other agents), guaranteeing that at least some information gets transmitted.
\end{enumerate}

The uniformity criterion corresponds to truthfulness: an untruthful agent would be one that uses a different mapping to the one everyone has agreed on. If everyone agrees that \texttt{0} means `bad standing', an agent that uses \texttt{0} to map to \emph{good} standing is untruthful.

\subsubsection{Cooperation}
Cooperation is paying a cost, $\gamma$, for the benefit, $\beta$, of someone else, stipulating that $\beta > \gamma$, that is, cooperation produces a net benefit for the aggregate of agents (there are of course interesting cases, beyond our scope, where $\gamma > \beta$). If an interaction consists of two agents coming together and deciding whether to cooperate, we obtain the payoff matrix in Table~\ref{prisoners-dilemma}. This is a prisoner's dilemma, because $\beta > \beta - \gamma > 0 > -\gamma$. Agents in our model will interact repeatedly in prisoner's dilemmas, each round with a new partner.

\begin{center}
\begin{tabular}{c c c c}
 & \multicolumn{3}{c}{Agent B} \\
 & & \texttt{d} & \texttt{c} \\
 \cline{3-4}
 \multirow{2}{4em}{Agent A} & \texttt{d} & \multicolumn{1}{|c}{$0$} & \multicolumn{1}{|c|}{$\beta$}  \\ 
 \cline{3-4}
 & \texttt{c} & \multicolumn{1}{|c}{$-\gamma$} & \multicolumn{1}{|c|}{$\beta - \gamma$} \\  
 \cline{3-4}
\end{tabular}
\captionof{table}{\textbf{Agent A's payoff in the prisoner's dilemma.} \texttt{d} stands for defect and \texttt{c} stands for cooperate.}\label{prisoners-dilemma}
\end{center}

\subsubsection{Stability} \label{section:model-stability}
Strategy $A$ is stable if no other strategy can invade it, that is, no other strategy can proliferate in a population of 100\% $A$-type agents (when $A$ is said to `predominate').

To make this precise, let $\mathbbm R_{x, 1}(y, z)$ be the expected payoff of strategy $y$ against strategy $z$ when a randomly selected agent has strategy $x$ with probability $1$; our requirement of stability can be stated as:
% TODO: this notation requires that x be either y or z if specification of its probability is to fully specify the population proportions.
\begin{equation}
\mathbbm R_{A, 1}(A, A) > \mathbbm R_{A, 1}(B, A)
\end{equation}

In words: $A$s do better against $A$s than $B$s do, when $A$ is predominant in the population. Since interactions will be with an agent of type $A$ with probability $1$, this is equivalent to the requirement that $A$ strictly outperforms $B$. We must take expectations since payoffs are random.

This condition is stronger than Maynard Smith's \cite{smith-the-logic-of-animal-conflict-1973}: we require that $A$ strictly outperforms any potential invader. Meeting this stronger requirement means that $A$ will not only be stable against invasions of a single strategy, but also of a combination of strategies. For instance, Maynard Smith's conditions allow for the possibility for a third, spoiler strategy $C$, against which $B$ does so well that its relative disadvantage is negated. Our definition of stability is robust against third `spoiler' strategies because if $A$ \emph{strictly} outperforms all other strategies from the start, there is no possibility that a third strategy would increase its population enough to become a spoiler.

\subsection{The model (Model 1)} \label{section:model1-the-model}
\begin{center}
    \begin{figure}[htp]
    \centering \centerline{
    \includegraphics[width=10cm]
    {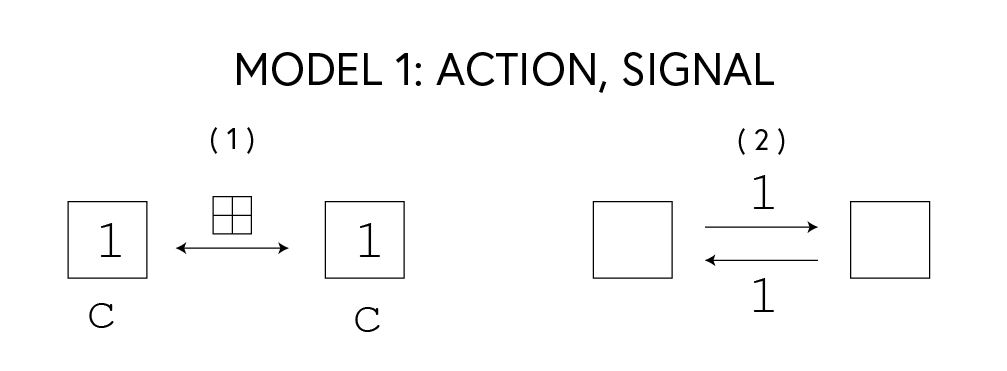}}
    \caption{\textbf{The focal strategy: stern discriminators interacting in Model 1.} In stage (1), the agents interact in a prisoner's dilemma, and can view each other's tags from the previous round. Agents then send signals about each other in step (2), producing new tags for the next round. Pictured is the behavior of the stern discriminator, which cooperates with \texttt{1}s and defects with \texttt{0}s. All agents who follow the this action protocol receive a \texttt{1}-tag from stern discriminators, and get cooperated with in the following round. Others receive a \texttt{0}-tag, and get defected with. Thus the strategy rewards the behaviors it prescribes; it is an \emph{aligned} norm.}
    \label{fig:model-1}
\end{figure}
\end{center}

% TODO: clarify that reproduction happens after the round of infinte interactions
The model unfolds in an infinite number of rounds in which agents interact in prisoner's dilemmas.\footnote{Following the infinite number of rounds, there is an implicit reproductive step that we do not directly model such that one's representation in the next generation is directly proportional to one's success (i.e., payoffs).} Agents also signal by `tagging' each other with a single bit of information -- \texttt{0} or \texttt{1}. Tags indirectly affect an agent's payoff because their partners might act differentially based on their tag. For simplicity, agents can only have one tag at a time, so whenever they are tagged, their previous tag is overwritten. A round consists of two steps:

\subparagraph{Model 1 description}
\begin{enumerate}
    \item Everyone pairs up with a partner at random, and makes a choice about whether to cooperate or defect in a prisoner's dilemma. An agent's choice may depend on their partner's tag from the previous round (Figure \ref{fig:model-1}(1)).
    \item \label{changed-step} Everyone signals a new `tag' for their partner. They can use whatever mapping they desire from the information they have (for instance, their partner's action, their action, their partner's old tag, their old tag, etc.) to the available symbols (\texttt{1} and \texttt{0}). The mapping can even be random -- but we assume agents tag each other simultaneously and thus do not know how the other agent has tagged them. This process can be imagined as your partner writing a \texttt{0} or \texttt{1} on your forehead for your next partner to see, as the signal need only be known to one's next partner  (Figure \ref{fig:model-1}(2)).
\end{enumerate}

The major difference between this indirect reciprocity model and others is in step (2) -- most, with the exception of \cite{smead-indirect-reciprocity-and-the-evolution-of-moral-signals-2010}, `lock' the communication system, pre-ordaining that agents must, for example, signal \texttt{0} about defectors and \texttt{1} about cooperators \cite{leimar-evolution-of-cooperation-through-indirect-reciprocity-2000, nowak-evolution-of-indirect-reciprocity-2005, nowak-evolution-of-indirect-reciprocity-by-image-scoring-1998, ohtsuki-the-leading-eight-social-norms-that-can-maintain-cooperation-by-indirect-reciprocity-2005, pacheco-stern-judging-a-simple-successful-norm-that-promotes-cooperation-under-indirect-reciprocity-2006, yamamoto-a-norm-knockout-method-on-indirect-reciprocity-to-reveal-indispensable-norms-2017}. However, we are interested in the conditions required to stabilize a particular communication system, and must therefore allow it to vary.

\subsection{Strategies} \label{section:model1-strategies}
To answer our question, we will seek to find a stable strategy (that includes both a prescription for how to act \emph{and} how to signal) that, when followed by everyone in the population, leads to a \emph{stable} state of (1) high cooperation and (2) effective communication. We will call this the focal strategy.

We want a relatively \emph{parsimonious} strategy that is stable; for this reason, we keep the set of example strategies in the coming sections simple. We will show, however, that the focal strategy is stable against \emph{all} potential invaders, not just invaders from the restricted set we present in the coming sections.

\subsubsection{Action}\label{section:model1-strategies-action}

% RESOLVED should we note here that 0 and 1 have arbitrary meaning? Perhaps we need to sprinkle in "without cost of generality statements" here and there?
One can imagine many possible action strategies. For example, the set of action strategies that take into account only the tag of the actor's partner are (1) cooperator (cooperate with everyone), (2) defector (defect with everyone), (3) discriminator (defect with \texttt{0}s, cooperate with \texttt{1}s), and (4) reverse-discriminator (defect with \texttt{1}s, cooperate with \texttt{0}s). These possibilities are shown in Table~\ref{strategies}. Note that (3) and (4) could just as easily have swapped names, because the symbols \texttt{0} and \texttt{1} are, a priori, meaningless. Furthermore, one could imagine action strategies with many more possible inputs, such as one's own tag or one's previous action. In the end, we will find a strategy stable against any other strategy, including these more complex ones.

\begin{center}
\begin{tabular}{ c|c|c|c}
\multicolumn{4}{c}{Action Strategies} \\
Name & ID & Action for \texttt{0}s & Action for \texttt{1}s\\
\hline
Defector & \texttt{dd} & \texttt{d} & \texttt{d} \\ 
Discriminator & \texttt{dc} & \texttt{d} & \texttt{c} \\ 
Reverse-discriminator & \texttt{cd} & \texttt{c} & \texttt{d} \\ 
Cooperator & \texttt{cc} & \texttt{c} & \texttt{c} \\ 
\end{tabular}
\captionof{table}{\textbf{Some possible action strategies} Each row represents a possible action strategy. The first and second bit in the ID specify how to act towards a partner with tags of \texttt{0} and \texttt{1}, respectively.} \label{strategies}
\end{center}

\subsubsection{Signal}\label{section:model1-strategies-signal}
After acting, agents may signal about their partner by giving them a one-bit tag. An agent's tag is observed by their subsequent partner (unless it is overwritten before then).

The signaling scheme we consider as a candidate for a stable state takes into account the action of the agent being tagged and the tag of their partner (who also happens to be the signaler, since agents tag their partners). Thus, there are four situations to be distinguished: cooperating with a \texttt{0}, cooperating with a \texttt{1}, defecting with a \texttt{0}, and defecting with a \texttt{1}. Because this class of signaling strategies takes into account both the actor's present action and their partner's action from one time step back (preserved by their tag), they are sometimes called `second order norms'.

There are 16 possible signaling strategies in this class, as there are two possible signals for each of these four outcomes. We will sometimes denote a signal strategy as a string of four binary digits (obtained by reading off the columns in Table~\ref{table:signal-strategies} vertically). For instance, \texttt{1001} corresponds to the stern judging signal strategy.

We will also consider the possibility that invading strategies might not signal at all in certain situations -- this will pose no problems to our stability results.

\begin{center}
\begin{tabular}{ c c|c c c}
& & \multicolumn{3}{c}{Signal of actor's new tag} \\
Partner's tag & Actor's action & Image Scoring & Standing & Stern Judging \\
\hline
\texttt{0}& \texttt{d(efect)} & \texttt{0} & \texttt{1} & \texttt{1} \\ 
\texttt{0} & \texttt{c(ooperate)} & \texttt{1} & \texttt{1} & \texttt{0} \\ 
\texttt{1} & \texttt{d} & \texttt{0} & \texttt{0} & \texttt{0} \\ 
\texttt{1} & \texttt{c} & \texttt{1} & \texttt{1} & \texttt{1} \\ 

\end{tabular}
\captionof{table}{\textbf{Three examples of signal strategies.} There are 16 of these, which would produce too many columns to reproduce here, but here are three possibilities. This class of signal strategies can take into account the tag of the actor's partner (the person affected by the action). Image-scoring does not care about the partner's tag, while standing and stern-judging do.} \label{table:signal-strategies}
\end{center}

\subsubsection{The focal strategy} \label{section:model1-strategy-focal-strategy}

Now it is time to select a candidate focal strategy (the strategy whose stability we will analyze) from the set we considered above. Consider the stern-judging discriminator:
\begin{enumerate}
\item Action: cooperates with \texttt{1}s, defects with \texttt{0}s
\item Signaling: tags cooperation with \texttt{1}-agents and defection with \texttt{0}-agents with a \texttt{1}, and cooperation with \texttt{0}s and defection with \texttt{1}s with \texttt{0}.
\end{enumerate}
We will henceforth call these agents `stern discriminators', or use the strategy string notation \texttt{dc1001}. Why is this strategy particularly promising? Partially because we have the benefit of hindsight. But there are good reasons to believe that stern discriminators may successfully fend off incursions of other strategies: the signal strategy tags everyone who follows the action strategy's dictates with a \texttt{1}, and the action strategy cooperates with \texttt{1}s, so anyone who follows the strategy gets cooperated with. Inversely, anyone who doesn't follow the action strategy gets defected with.

In other words, the strategy is admirably self-consistent -- it rewards all of the behaviors it prescribes and punishes those that it doesn't. If everyone follows it, it is an \emph{aligned norm}. The intuition, then, is that any alternative will do worse in a world dominated by stern discriminators because they will punished more and rewarded less. 

In the following section we will see if this intuition bears out, by analytically determining the stability of the stern discriminators. To make the mathematics tractable, we will make two simplifying assumptions: first, we will assume an infinite population; second, as already described in Section~\ref{section:model1-the-model}, we will assume infinite rounds per generation, so that agents' actual payoffs converge to their expected payoffs (for further justification, see \ref{appendix-approximation-justifications}).
% TODO: circle back to appropriately reference the appendix

\subsection{Analysis}
The analysis will proceed by checking if the stern discriminator state is 
\begin{enumerate}
    \item \emph{cooperative} (where we define `cooperative' as a world in which at most $\epsilon$-fraction of agents defect, where $\epsilon > 0$ but can be made arbitrarily small),
    \item \emph{effectively communicative}, and
    \item \emph{stable}. 
\end{enumerate}

\subsubsection{Cooperative and communicative: check}
First, cooperative. Let us suppose agents start with an arbitrary assignment of tags (could be all \texttt{0}, all \texttt{1}, or any mix). Then, in the first round, stern discriminators will cooperate with \texttt{0}s and defect with \texttt{1}s. Because all who follow the stern discriminator strategy receive a tag of \texttt{1}, everyone will receive a \texttt{1}-tag to start the second round, and thus, everyone will cooperate with their partners in the second round. Everyone will thus receive a \texttt{1}-tag \emph{again} and the cycle will continue -- so all rounds except the first are fully cooperative (Figure~\ref{fig:progression-of-rounds}(A)).

Clearly the state consisting entirely of stern discriminators is effectively communicative, since all agents abide by a separating, uniform mapping (as specified by the stern judging norm), and never withhold information (see Section~\ref{section:effective-communication}).

\subsubsection{Stability: fail}
It remains to check if the state is stable (spoiler: it won't be). In this section, we will identify what the instabilities of this state are, paving the way for the additional mechanisms we propose in the next section.

It turns out that many alternative strategies can invade this model. Two instabilities (that is, vulnerabilities that allow alternative strategies to perform just as well) lie at the root of these invading strategies:
% TODO** : maybe a figure of the invading strategies 
\begin{center}
    \begin{figure}[htp]
    \centering \centerline{
    \includegraphics[width=16cm]
    {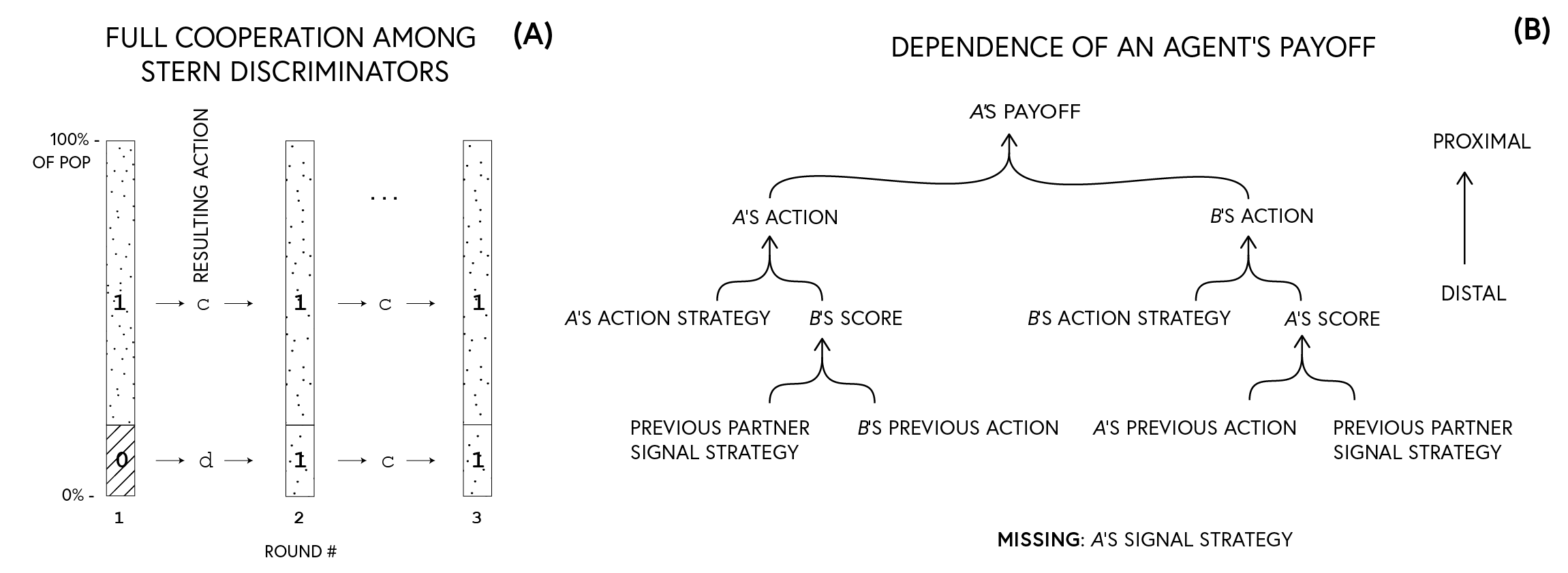}}
    \caption{
    \textbf{(A) Proportion of \texttt{1}-tagged agents.} In the first round, agents are arbitrarily assigned \texttt{1}- and \texttt{0}-tags. Stern discriminators cooperate with the \texttt{1}s and defect with the \texttt{0}s, and all receive a \texttt{1}-tag (according to the stern judging signal strategy) as a result. This leads to cooperation forever after.
    \newline
    \textbf{(B) Factors influencing an agent's present payoff.} Many factors from previous rounds determine an agent's payoff in the current round. However, none of them have to do with their own signaling strategy}
    \label{fig:progression-of-rounds}
\end{figure}
\end{center}
\subparagraph{Instabilities in Model 1}
\begin{enumerate}
    \item \textit{Unpunishability of language} - In the base model, there is no payoff difference between individuals with the same action strategy but different languages. This is because agents' payoffs depend on two things: their own actions and their partners' actions, neither of which depend on their own signal strategies (see Figure~\ref{fig:progression-of-rounds}(B)). Thus, given an action strategy that can perform as well as the stern discriminator, swapping out its signal strategy will not hamper its ability to invade (e.g. a `pushover' discriminator that signals \texttt{1} about everyone does just as well as a stern discriminator).
    \item \textit{Unexpressed traits} - We know that in the stern discriminator world, everyone cooperates, and therefore everyone receives a tag of \texttt{1}. Therefore, no one ever must decide how to act against an agent with a tag of \texttt{0}, and no one must decide how to tag an agent who has defected (assuming that agents interact with mutants with negligible frequency). This means that the part of the action strategy that prescribes how to act against \texttt{0}s, and the part of the signal strategy that prescribes how to signal about defectors (along with the part that prescribes how to signal about cooperation against a \texttt{0}-tagged agent) remain latent. Thus, for example, unconditional cooperators are indistinguishable from discriminators.
\end{enumerate}
Now referring to our strategy-string nomenclature, unexpressed traits imply that there are four strategy clusters, each of whose members have identical payoffs (where a dash indicates that the `slot' can be filled with either of its two possibilities): \texttt{-d---0}, \texttt{-d---1}, \texttt{-c---0}, and \texttt{-c---1}. Payoffs within each cluster are identical because they only differ on traits that never get expressed. And because the last two clusters differ visibly only on their signal strategies, we obtain that the last two clusters have identical payoffs. We conclude that any member of the last two clusters can invade.

Still, the stern discriminators can at least fend off the defecting clusters: the \texttt{-d---1} and \texttt{-d---0}. This is because their defections will provoke stern discriminators to \texttt{0}-tag them (since defection against \texttt{1}-tagged agents does not conform to the stern-discriminator norm). This will lead their subsequent partners to defect against them, generating for them a payoff of zero (compared to $\beta - \gamma$ for stern discriminators).

\section{Model 2: next!}\label{section:model2}
In this section, we address the two instabilities of the base model with two modifications. First, to address the irrelevance of language, we add a mechanism, called meta-signaling, to the model. This allows agents to signal not only about each other's actions but also about each other's signals, thereby making language relevant by folding the truthfulness of an agent into their reputation.
\begin{center}
 \begin{tabular}{r p{8cm}}
        \textbf{Instability} & \textbf{Remedy}  \\
        \hline
        Unpunishability of language & \emph{Meta-signaling} tags agents based on their signal \\[0.6em]
        \hline
        Unexpressed traits & \emph{Error} creates diversity\\
    \end{tabular}
    \captionof{table}{\textbf{Instabilities in Model 1, their remedies in Model 2}. Observers who can tag agents based on their signal are called \emph{meta-signalers}. Their presence means that each agent has a probability $p$ of receiving a tag based on their signal (even meta-signalers themselves) which makes signal relevant to their own payoff. The addition of \emph{error} means that agents deviate from their prescribed strategy with probabilities $\epsilon$ (in the case of action) or $\delta$ (in the case of signal). Error introduces heterogeneity required to bring all traits into expression at least some of the time.}\label{table:instabilities-and-remedies}
\end{center}

But meta-signaling alone is not enough. Even if agents can signal about each other's truthfulness, there are still no defections, and no agents with a \texttt{0}-mark, so any components of the strategy relevant to those cases will remain latent. To remedy this second instability, we add error: agents deviate from their prescribed strategies with some small probability ($\epsilon$ for actions, $\delta$ for signals). This introduces the diversity required to bring out latent traits. We summarize the instabilities and their remedies in Table~\ref{table:instabilities-and-remedies}.

% RESOLVED BELOW it is not 100% clear from the preceding description which error rate applies to the normal tagging as per Model 1.

\subparagraph{Model 2 description}

\begin{enumerate}
\item In each round, agents are assigned either to the role of \emph{actor} (with probability $1 - p$) or \emph{observer} (with probability $p$).
\item Actors are paired randomly. 
\item Observers are assigned randomly to another agent -- note that the agent being observed can be either an actor or another observer. Thus, assuming an infinite population, and that each agent has a maximum of a single observer, one's probability of being observed in any given round is $p$ (we will show in Section~\ref{section:discussion-relaxation} that the single-observer requirement is not strictly necessary).
% RESOLVED this implies that there is always some probability of a long stretch with no observations, which matters for the payoff calculations
\item Actors interact in a prisoner's dilemma, making cooperation and defection choices based on what their strategies prescribe (Figure~\ref{fig:model-2}(1)), with error rate $\epsilon$ (see Section~\ref{section:model2-strategies-and-error}).
% RESOLVED we might be precise here about what the error entails; for now, I have added "see below"
\item Actors signal by tagging each other -- either a \texttt{0} or a \texttt{1} -- based on their signaling strategies (with error rate $\delta$). An agent's tag will be observable by whoever next interacts with them (as long as it is not overwritten before then), in addition to the observers of the interaction (more on this in Section~\ref{section:model2-strat-meta-signaling}). Depicted in Figure~\ref{fig:model-2}(2).
\item Observers `meta-signal' by determining whether they agree with their observee's signal, and they tag their observee according to their meta-signaling strategy (also with error rate $\delta$), overwriting the observee's pre-existing tag (Figure~\ref{fig:model-2}(2)). 
\end{enumerate}

\subsection{Strategies and error} \label{section:model2-strategies-and-error}
% RESOLVED the following line clarifies the preceding RESOLVED circle back to make sure no confusion will result.
Action and signaling work the same in this new model, with the exception that agents make errors: that is, agents follow the prescriptions of the action and signal strategies only with probability $1 - \epsilon$ and $1 - \delta$, respectively -- otherwise, they do the opposite. Apart from the addition of error, observers make the difference between Model 2 and Model 1, so we will first describe their behavior in more detail, and then define an extended focal strategy that includes a meta-signaling strategy (the stern discriminators have a defined action and signal strategy, but we've not yet said how they should meta-signal).

\subsubsection{Observers (meta-signalers)}\label{section:model2-strat-meta-signaling}
\begin{center}
    \begin{figure}[htp]
    \centering \centerline{
    \includegraphics[width=10cm]
    {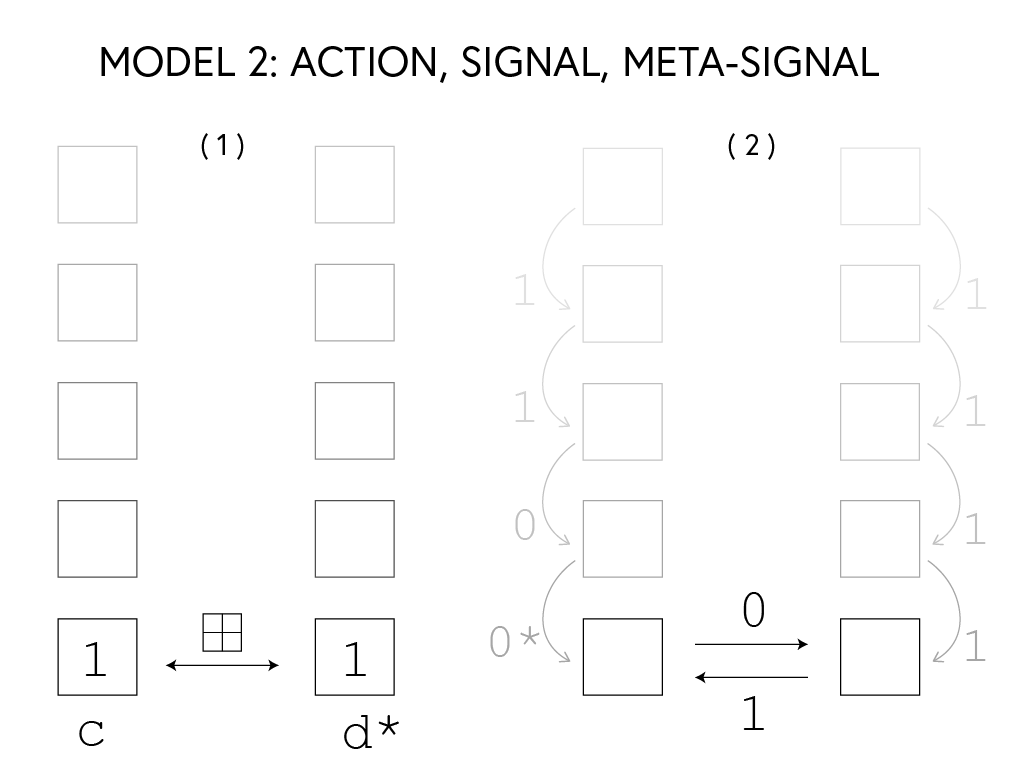}}
    \caption{\textbf{Action and signaling in Model 2 with stern discriminators.} In stage (1), agents make decisions on how to act. The asterisk indicates that the defection from the right-hand agent is an action that doesn't conform to the stern discriminator strategy. Every agent is observed with probability $p$ (the decreasing opacity of the observers represents the geometric fall-off in the probability of their presence). Then, in stage (2), the actors tag each other, with the non-conforming actor getting a \texttt{0}-tag. Agents also receive a tag from their observer (if they have one), based on whether their observer agrees or disagrees with their tag (e.g. the second agent from the bottom on the left sends a non-conforming signal, hence in turn receives a \texttt{0}-tag.)}
    \label{fig:model-2}
\end{figure}
\end{center}

Every individual, whether actor or observer, has someone watching them with probability $p$. There can be a chain of observers: perhaps a first-level observer is observing one of the actors, and some second-level observer is observing the first-level observer (see Figure~\ref{fig:model-2}). If $p$ is small, however, these chains are unlikely to be very long -- their expected length is $\frac{1}{1 - p} - 1$, derived from the waiting time for an event (not having an observer) that has probability $1 - p$.
% RESOLVED explain in the discussion how appealing it is to have the snake biting its own tail in this manner (e.g., it gets around the problem of infinitely recursing and pointing out that, at each stage, people may not behave according to the presumed assumptions at that level of recursion).

We assume the observer can see the chain of events (actions, signals, and meta-signals) before it. Using this information, observers can `meta-signal' according to some strategy (named so because they are signaling about someone's signal). This meta-signal then overwrites the observee's previous tag.

% RESOLVED Since, there are other ways of specifying the metasignaling strategy that do not depend on the agree/disagree distinction (acknowledged below)), we should word the following sentence to make sure we do not imply otherwise. e.g., One way to specify metasignaling strategies is to consider ether the observer agrees or disagrees with...
One way to specify meta-signaling strategies is to consider whether an observer would have signaled in the same way as the agent they are observing (they agree) or in a different way (they disagree). If the signaler fails to signal when the observer would have, this counts as a `disagree'. An agent follows the so-called `separator' meta-signaling strategy if they signal \texttt{1} when they agree and \texttt{0} when they disagree.

% RESOLVED I think we need more details here on what agree with and disagree with entail.

Of course, one could imagine a wealth of much more complex meta-signaling strategies; we will, however, show the new focal strategy to be stable against all such alternatives.

\subsubsection{The new focal strategy}
In the spirit of the rest of the stern discriminator strategy, we say that the new focal strategy uses the `separating' meta-signaling strategy: \texttt{0} if they disagree, \texttt{1} if they agree.

This selection maintains the property that the strategy rewards the behaviors it prescribes: anyone who is observed following the prescribed signaling and meta-signaling strategies gets a \texttt{1}-tag from the observers, leading to cooperation in the subsequent round. By contrast, anyone who is observed failing to follow the signaling and meta-signaling strategies gets a \texttt{0}-tag and a subsequent defection.

\subsection{Analysis}
In this section, we will test to see if model 2 is cooperative, effectively communicative, and stable (it will be). We assume a population of mostly stern discriminators (future work could address arbitrary mixes of strategies).

\subsubsection{Cooperative and communicative: check}
The state is clearly effectively communicative since all agents abide by a separating mapping from meanings to symbols, never withholding information. To see that it is cooperative, we show in \ref{appendix-model2-cooperative} that while it is true that $\epsilon > 0$ and $\delta > 0$ imply that there are a nonzero portion of defections, that portion can be made arbitrarily small by shrinking $\epsilon$ and $\delta$.

\subsubsection{Stability: check}
 Learning from the failures of Model 1, for stability we must check that
\begin{enumerate}
\item \emph{Any} strategy that deviates from the stern discriminators, in a world dominated by stern discriminators, does strictly worse.
\item No latent traits exist that will cause invading strategies to be indistinguishable from stern discriminators.
\end{enumerate}

\paragraph{Any alternative does worse} \label{section:any-alt-worse}

\begin{center}
    \begin{figure}[htp]
    \centering \centerline{
    \includegraphics[width=18cm]
    {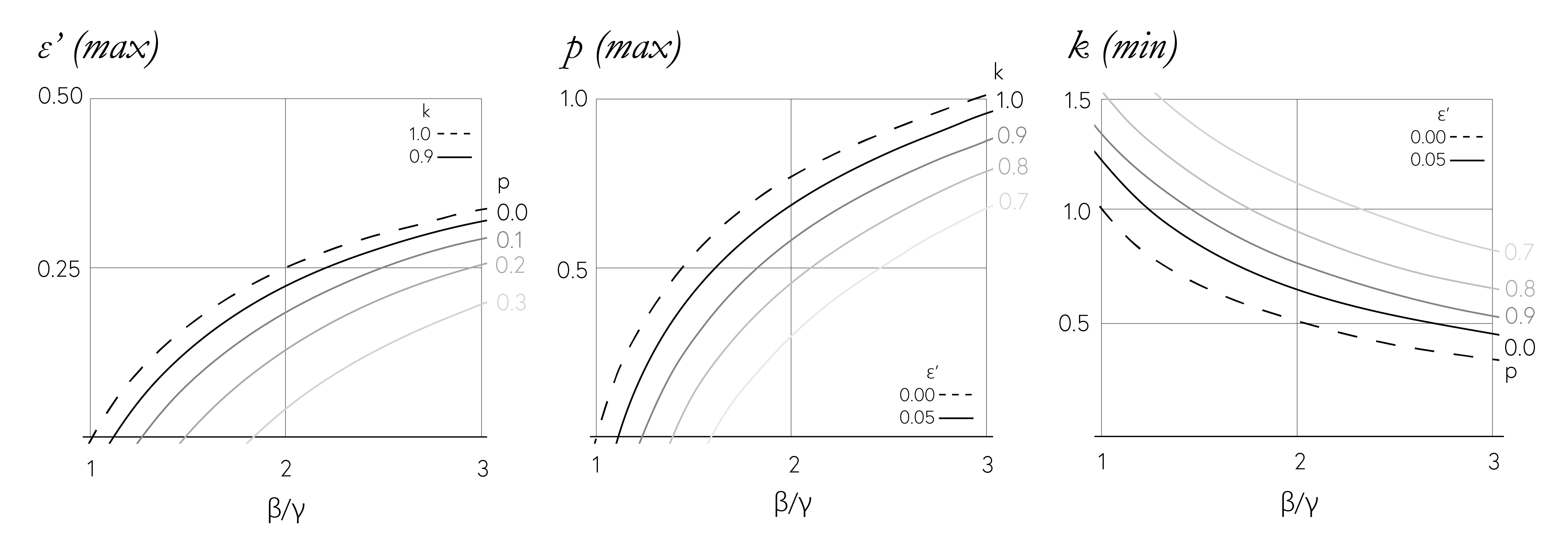}}
    \caption{\textbf{Maximum (minimum) values of $\epsilon'$, $p$, and $k$ under (over) which any alternative strategy is costly.} Inequalities are strict, so values on the lines do not give stable cooperative regimes. By setting the other two (non-y) variables to values arbitrarily close to the dotted line, the y-variable can approach, but never attain, the dotted value.}
    \label{fig:bounds}
\end{figure}
\end{center}

We start with the first step. Whenever an agent makes any decision -- whether on how to signal or how to act -- there are three possible consequences of that decision:
\begin{enumerate}
    \item \emph{first-order costs} - immediate costs paid for a decision (for instance, paying the cost $\gamma$ of cooperation)
    \item \emph{second-order costs} - costs borne for a decision in the subsequent action round (for instance, their subsequent partner defects with them for having a \texttt{0}-tag)
    \item \emph{effects on future scenarios} - the present decisions of agents may affect the conditions surrounding their future decisions, potentially affecting payoffs.
\end{enumerate}

% TODO: Victor to check that the following sentence is sufficiently correct.
These costs are the changes in payoffs relative to following the stern discriminator strategy. Thus, to show that deviating is always costly, we need to show that 

\begin{enumerate}[label=\roman*.]
    \item \emph{Given} a particular decision-making scenario in a world dominated by stern discriminators, deviating from the stern discriminator strategy is costly (the sum of the first-order (1) and second-order (2) costs is positive), and
    \item In no case does deviating in a particular scenario produce future scenarios with higher expected payoffs (3).
\end{enumerate} 

The combination of these two facts implies that deviating is always, on net, costly.

The intuition behind part (i) of the proof is that the first-order benefits of deviating are at most around $\gamma$ (since you can save on the cost of cooperating, potentially, by deviating), but the second-order costs can always exceed this since they are proportional to $\beta$ (stern discriminators will withhold cooperation from those who deviate, costing a deviant $\beta$), and $\beta$ is greater than $\gamma$. For part (ii), we will show  that deviating often has no effect on future decision-making scenarios, but even when it does, it cannot increase one's payoff.

We will conclude that there are a set of parameters $p$, $\epsilon$, $k$, and $\delta$ for which \emph{any alternative strategy}, pure or mixed, does worse.

Let us be precise with these bounds (shown in Figure~\ref{fig:bounds}). To meet our stability criterion, we need our effective error rate (defined as $\epsilon' := \epsilon + \delta - 2\epsilon\delta$) to satisfy

\begin{equation}
0 < \epsilon' < \dfrac{1}{2} - \dfrac{\gamma(1 - p(1 - p)k)}{2\beta(1 - p)^2k}
\end{equation}

The higher the benefit is relative to the cost, the closer the error rate can be to $\frac{1}{2}$.

Then we define $r = \frac{(1 - 2\epsilon')\beta}{\gamma}$, so named because it is the ratio between the expected second-order benefit of cooperating and the first-order cost (the benefit is not quite $\beta$ because of potential error). The probability of being an observer, $p$, must satisfy
\begin{equation}
0 < p < \dfrac{k(2r - 1) - \sqrt{k^2 - 4k + 4kr}}{2k(r - 1)}
\end{equation}

Finally, the discounting factor, $k$, must satisfy

\begin{equation}
    k > \dfrac{\gamma}{(1 - p)^2(1 - 2\epsilon')\beta + \gamma(1 - p)p}
\end{equation}
If $\epsilon' = p = 0$, this would collapse down to $k\beta > \gamma$, which makes sense as the discounted benefit would need to exceed the immediate cost (this relation is represented by the dotted line in the left panel of Figure~\ref{fig:bounds}). The added complexity in the denominator stems from the fact that not every defection is punished.

Some readers may choose to jump straight to Section~\ref{section:all-traits-expresed}, skipping the proof. 

\subparagraph{In a given scenario, deviation is costly}
In this part, we show that in a world of stern discriminators, the total cost of deviating in a particular scenario (the sum of first and second order costs) is  positive. We define the cost of doing $x$ as the \emph{payoff difference} between doing $x$ and doing what a stern discriminator would have done.

First-order costs are borne immediately, and second-order, if ever, in the subsequent action round. Thus, we can calculate the expected costs of an agent's choice in action round $n$ simply by looking at their payoffs in \emph{action} rounds $n$ and $n + 1$.

\textbf{First-order costs (round $n$)} There is no immediate cost or benefit to signaling in any particular way. All potential signaling costs are second-order. Action, however, is a different story: there are immediate payoff differences between acting conformingly and not, because cooperation is always more costly than defection.

% QUESTION I think we should make it even more clear that the payoffs being calculated are for the focal strategy, not an arbitrary strategy -payoffs are for any deviation from the stern discrim

These costs are quite simple. There are two ways of acting non-conformingly: cooperating with a \texttt{0} or defecting with a \texttt{1}. Cooperating with a \texttt{0} comes with an additional cooperation cost $\gamma$. On the other hand, defecting with a \texttt{1} produces some upfront benefit (negative cost), as the agent avoids paying $\gamma$. Thus, defining $F_x$ as the first-order cost of deviating when doing $x$ (e.g. $F_\text{def}$ is the first-order cost of deviating by defecting), we have

\begin{equation}
\begin{aligned}
F_\text{sig} &= 0 \\
F_\text{meta-sig} &= 0 \\
F_\text{def} &= -\gamma \\
F_\text{coop} &= \gamma
\end{aligned}
\end{equation}

\textbf{Second-order costs (round $n + 1$)} Now we calculate the second-order costs of deviating from the stern discriminator strategy. Conveniently, this looks pretty similar for action, signal, and meta-signal deviations. To pay a second-order cost for deviating, three things must happen.
\begin{enumerate}
    \item[\emph{Tag}] The deviant must receive a tag for the behavior in which they strayed from the norm; for instance, if the agent deviated in action, the agent should receive a tag for their action and not their signal
    \item[\emph{Stick}] That tag must stick until the subsequent action round, so that their partner has an opportunity to punish them for it
    \item[\emph{Norm}] Both the agent tagging the deviant and the agent subsequently acting with the deviant must not err, because if they do, they will fail to punish the deviation -- that is, the norm must be effectively followed.
\end{enumerate}

% RESOLVED I think we need to define cost. The way it's used in the proof, it seems always to be the difference in expected payout comparing two alternative strategies.
% RESOLVED one thing that we could make clearer about the proof is that it assumes throughout that the focal strategy dominates in the population
We will start by assuming all three of these conditions, then gradually roll back to get the actual expected second-order cost of deviation. First, assuming all three conditions, we have the second-order cost of deviating,
\begin{align}
    S_\text{tag, stick, norm} = \beta,
\end{align}
because if all the conditions are satisfied, the deviant's subsequent partner will defect, costing them the benefit from cooperation $\beta$ (there is technically a discount factor here, but we leave it out since we do not yet know how many rounds the agent waited between action rounds).

If we then roll back the assumption that the norm was effectively followed (because someone erred) in the pipeline, what is the expected cost of deviation? Where $S_\text{tag, stick}$ is this quantity, we have

% I'm mildly worried that we have to add a term for the (small) proportion of non-focal individuals. Even though it's small, so are epsilon and delta, and perhaps there's an interaction that changes things.
\begin{align}
S_\text{tag, stick} &= (1 - \epsilon)(1 - \delta)\beta + \epsilon\delta\beta + \epsilon(1 - \delta)(-\beta) + (1 - \epsilon)\delta(-\beta) \label{equation:r1-r0-difference}\\
&= (1 - 2(\epsilon + \delta - 2\epsilon\delta))\beta \nonumber \\
&:= (1 - 2\epsilon')\beta. \label{equation:eff-error}
\end{align}

\noindent In Equation~\ref{equation:r1-r0-difference}  we distinguish between four scenarios: (1) neither the deviant's tagger nor their current partner erred, (2) both erred, or (3, 4) one erred and not the other. If neither the deviant's tagger nor their subsequent action partner errs, or if they both err, the deviant forgoes the benefit of cooperation (paying cost $\beta$). However, if one of the two makes an error, the deviant's subsequent partner actually cooperates, giving them a benefit of $\beta$. The equation calculates the cost, so benefits are negative and costs are positive.

Notice that we defined the effective error rate in Equation~\ref{equation:eff-error}, $\epsilon' = \epsilon + \delta - 2\epsilon\delta$. Here is why. Intuitively, an error in this scenario is when someone who deviated doesn't get the response that one would expect. This happens either if the deviant's signaler makes an error and the subsequent partner acts faithfully on the erroneous signal, or if the signaler is accurate but the subsequent partner makes an error. These scenarios respectively have probabilities $\delta(1 - \epsilon)$ and $\epsilon(1 - \delta)$, and adding them up we obtain $\epsilon + \delta - 2\epsilon\delta$.

But what if we roll back the sticking assumption? This requires us to account for two things. First, the tag might be overwritten between the deviation in behavior and the subsequent action round. And second, the longer we wait between the deviation and the subsequent action round, the more the resulting cost will be discounted. 

\begin{align}
S_\text{tag} &= (1 - p)kS_\text{tag, stick} + p(1 - p)kS_\text{tag} \label{equation:ctag-recursive} \\
&= \dfrac{(1 - p)k}{1 - p(1-p)k}S_\text{tag, stick} \nonumber
\end{align}
The first term in Equation~\ref{equation:ctag-recursive} corresponds to the scenario where the subsequent round is an action round, with probability $1 - p$, and therefore the agent pays $S_\text{tag, stick}$, but discounted by a factor $k$ since a single round has passed. The second term corresponds to the case where the agent is an observer in the subsequent round (which happens with probability $p$) but is herself unobserved (which happens with probability $(1 - p)$) and thus the tag persists for another round, and we are back to the initial question: what is the cost assuming you have been tagged for a deviation (of course, it we need to discount by a factor $k$, since in this case a round has passed by).

So we've obtained $S_\text{tag}$, which is the second-order cost of deviating given that you are being tagged by someone on that behavior. Note that $S_\text{tag} > 0$ when $\epsilon' < \frac{1}{2}$, which makes sense: as long as errors occur less often than not, agents must pay some expected second-order cost \emph{if they will be tagged based on their deviation}. But we are interested in $S_x$, the second-order costs of deviating when doing $x$, when an agent may or may not be tagged for it.

First, the signaling case. Agents are tagged based on their signal with probability $p$. So we obtain
\begin{align}
S_\text{sig} &= pS_\text{tag} \\
&= S_\text{meta-sig}
\end{align}

Meta-signaling in a non-conforming way has exactly the same second-order cost, $S_\text{meta-sig} = S_\text{sig}$, since the probability of being observed $p$, the probability of sticking $p_\text{stick}$, and the cost $S_\text{dev, tag}$ are all equal to their corresponding values in the regular signaling case.

In the action case, agents are tagged based on their action with probability $(1 - p)$ (in the event that their signal is not being observed). So this gives

\begin{align}
S_\text{coop} &= (1 - p)S_\text{tag} \\
&= S_\text{def}
\end{align}

\textbf{Total costs} We now would like to calculate the total cost, $C_x = F_x + S_x$ of deviating from the norm in each scenario, so that we can set bounds on the parameters that will allow for a stable stern discriminator state. We will start with the signaling case:

\begin{align}
C_\text{sig} = C_\text{meta-sig} &= F_\text{sig} +  S_\text{sig} \nonumber \\
&= p\dfrac{(1 - p)k}{1 - p(1-p)k}(1 - 2\epsilon')\beta \\
C_\text{def} &= F_\text{def} +  S_\text{def} \nonumber \\
&= (1 - p)\dfrac{(1 - p)k}{1 - p(1-p)k}(1 - 2\epsilon')\beta - \gamma \\
C_\text{coop} &= F_\text{coop} +  S_\text{coop} \nonumber \\
&= (1 - p)\dfrac{(1 - p)k}{1 - p(1-p)k}(1 - 2\epsilon')\beta + \gamma
\end{align}
For $C_\text{sig} > 0$, we just need $p > 0$, $\epsilon' < \frac{1}{2}$. And since $C_\text{coop} >  C_\text{def}$, if we can find the values for which $C_\text{def} > 0$ (that is, it is costly to deviate by defecting), then it will also be costly to deviate by cooperating. Solving for the values of $p$, $k$, and $\epsilon'$ that satisfy that inequality, we obtain the bounds we gave at the beginning of Section~\ref{section:any-alt-worse}.

\begin{center}
    \begin{figure}[htp]
    \centering \centerline{
    \includegraphics[width=8cm]
    {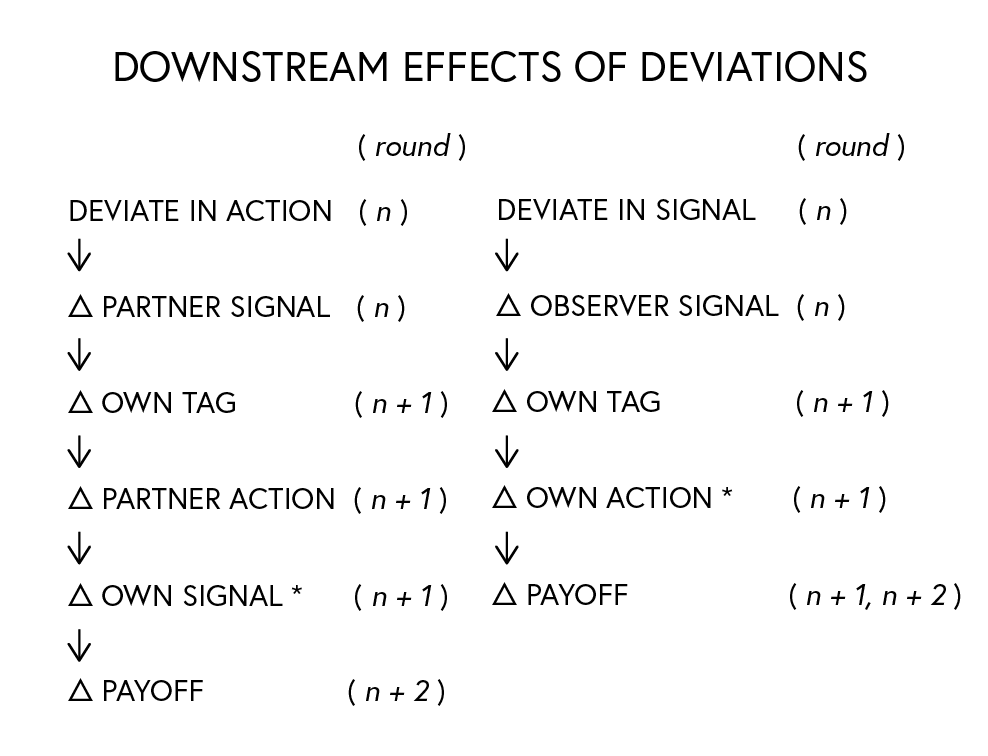}}
    \caption{\textbf{Ways that non-conforming decisions can create alternative downstream scenarios.} On the left side, an action deviation in round $n$ can lead to a difference in one's signal in round $n + 1$ which can lead to a payoff difference in round $n + 2$. Similar downstream effects can occur in signal deviations. However, we show that none of the downstream consequences of a deviation can actually benefit an agent (either they are neutral or detrimental).}
    \label{fig:downstream-effects}
\end{figure}
\end{center}

\subparagraph{Deviations cannot produce beneficial future scenarios}
We've shown that deviating in a given scenario is always costly, but can deviating produce future scenarios that are beneficial? Figure~\ref{fig:downstream-effects} shows ways that this might happen.

While deviant actions may alter the conditions surrounding future decisions, they can't improve future expected payoffs. Recall, for each type of decision, stern discriminators distinguish between the following world scenarios:
\begin{enumerate}[label=\alph*.]
    \item \emph{Action} - the tag of one's partner (two possible scenarios)
    \item \emph{Signal} - one's own tag and the action of one's partner (four possible scenarios)
    \item \emph{Meta-signal} - whether one agrees with the signal of the agent they are observing (two possible scenarios).
 \end{enumerate}

 We need to show that for each of these cases, a prior deviation cannot produce more beneficial subsequent scenarios. In part (a), it is true that interacting with agents with a \texttt{0}-tag is more beneficial because it means that an agent can conformingly defect, saving on the cost of cooperation. However, an agent's previous actions have no effect on the tag of one's partner, which are determined by the partner's previous interactions. So deviating cannot create more favorable \emph{action} scenarios.

 In part (b), it is possible for an agent's previous actions to influence both their own tag and their partner's action. However, conforming in both action and signal is the highest payoff action, since it leads to an expected subsequent payoff of $(1 - 2\epsilon')\beta$. Thus, deviating cannot lead to more beneficial signaling scenarios.

 In part (c), an agent's previous actions do not affect the signal of the agent they are observing (which is independent of the observer), so deviating in some prior round cannot create more favorable meta-signaling scenarios.

Thus, no matter the kind of decision, prior deviations cannot produce more beneficial future scenarios. This completes the proof.
 
\paragraph{All traits are expressed with error} \label{section:all-traits-expresed}
We've found parameter values for which it is always costly to deviate from the stern discriminator strategy. But could the homogeneity of the stern discriminator world restrict the set of possible world states in such a way that an invading strategy never deviates from stern discriminators on the \emph{restricted set} but is nonetheless different on the \emph{full set}? If this is the case, that strategy could invade by drift, causing vulnerabilities. We saw how, in Model 1, this vulnerability allowed any of the \texttt{-c---1} strategies to invade. However, this is no longer a problem in our current model, because

\textbf{Any bounded world state comes to pass,} due to \emph{error}. Let us consider an agent's world state to be some sequence of actions, signaling events, and tags that an agent has seen, either as an observer or an actor. The sequence is of bounded length (we assume agents have bounded memory).

Any arbitrary string of events of this form has nonzero probability, because a world state is a conjunction of finitely many agent decisions, all of which have nonzero probability due to error. Therefore, even a strategy that only deviates from stern discriminators in the most esoteric of world states would be caught deviating on occasion -- and therefore, would receive a lower payoff. 

\textbf{Variables defining world states are limited.} Of course, likelihoods for such esoteric events quickly become vanishingly small, which matters in practice since we don't actually have infinite rounds. However, also in practice, agents are under cognitive constraints that mean they will likely not be able to use a very long sequences as world states, and instead would probably make use of the information \emph{relevant} to the event. The world state for a given decision, then, would likely be determined by variables relevant to the situation at hand, such as one's own tag and action and one's partner's tag and action. When world states are determined by this smaller set of variables, any given one is very likely to occur at some point, even if we relax the infinite-round approximation.

\textbf{Taking advantage of rare world states is hard.} But suppose there is a world state that is extremely unlikely to occur once the infinite-round approximation is relaxed, and that strategy $B$ deviates in this world state but no others. This is a problem for the cooperative equilibrium not because Strategy $B$, which is indistinguishable from stern discriminators, can invade. The problem is that a third strategy, $C$, may invade by \emph{taking advantage of the scenario on which stern discriminators and Strategy $B$ differ}. 

Here is the classic example: in the model of indirect reciprocity with no error, unconditional cooperators can invade the world of image-scorers (those who defect with defectors and cooperate with cooperators), because no one in this world is a defector (and thus the situation in which image-scorers and cooperators differ is never realized).  This first wave of invaders doesn't perturb the cooperative equilibrium, but they create the opportunity for a third strategy, unconditional defectors, to enter without suffering any consequences (since cooperators cooperate with defectors). That is, the defectors take advantage of the situation on which image-scorers and cooperators differ,the scenario where someone defects, by \emph{creating} it: by themselves defecting.

In this model, it would be impossible for a lone invader to single handedly create such rare situations, which consist of unlikely combinations of one's own tag and action, one's partner's tag and action, and for meta-signalers, the chain of signals leading up to them, in addition to the tag of the agent they're observing. Out of these, the only levers an individual agent may pull to create an unlikely world state are their own tag and their own action. Unlike the unconditional defectors in the image scoring case, who can simply, by their own choices, produce the scenario they are able to take advantage of, these invaders must rely on serendipitous combinations of their own actions and errors of other agents (who will be of the predominant strategy). Thus, these agents cannot reliably create the scenario, making it extremely difficult to exploit.

In summary, there are three safeguards against the problem of unexpressed traits. First, any bounded world state is bound to occur eventually. Second, due to the structure of the interactions, minimal relevant information is available to an agent in any given round, paring down the plausible world states to a set whose members are all reasonably likely to occur. And third, even \emph{if} there was an extremely rare world state that doesn't occur once the infinite rounds assumption is relaxed, it would be very difficult to exploit because it relies on the confluence of deviations of many agents, not just the agent that would stand to benefit.

\section{Discussion}\label{section:discussion}
The three conditions under which communication and cooperation could remain stable, without assuming that any exogenous machinery was doing the work for us, were:

\begin{enumerate}
\item \textit{Aligned norms}, rewarding conforming action strategies
\item \textit{Meta-signaling}, rewarding conforming signaling strategies
\item \textit{Error}, ensuring that all components of the strategy specification (both signal and action) will find expression at least some of the time -- important because when certain components of the strategy remain unexpressed, there will be clusters of indistinguishable strategies, all of which will likely invade.
\end{enumerate}

We should note that these conditions apply equally to genetically and culturally transmitted strategies, and they are indifferent to whether alternative strategies invade an existing equilibrium by way of mutation, migration, or agents simply choosing to change their strategies. We will first examine these three stabilizers in more detail, and then go on to analyze the resulting equilibrium and its downstream consequences.

\subsection{The three stabilizers}
\subsubsection{Meta-signaling}

The concept of meta-signaling makes it clear \textit{why} it makes sense to study the evolution of communication in the context of cooperation. Research suggests that there must be some kind of pressure on the signaler for a uniform communication system to emerge \cite{oliphant-the-dilemma-of-saussurean-communication-1996}. This pressure may be intrinsic to the signal, or it may result from differential allocations of social benefits and costs to truth-tellers an deceivers. And this differential allocation --- whether through kin selection, reciprocity, punishment, or indirect reciprocity --- is \emph{also} how altruism becomes stable, at least conventionally. Because the machinery for such differential allocation \textit{is already present} for cooperation, it is plausible for it to have been co-opted for the purposes of communication. Meta-signaling co-opts the indirect reciprocity mechanism for the purposes of maintaining the communication system, but one can equally imagine a co-optation of other mechanisms, such as reciprocity (e.g., lying to those who have lied to you) and punishment (e.g., ostracizing those who miscommunicate) for the same purpose. 

% RESOLVED vertical and horizontal already are used for trasmission in cultural evolution. There's some chance here for confusion. I think we can leave the following paragraph as is, however, and see if reviewers are okay with our usage.

% Mike TODO you had a comment:

% missing word near the end -- should it be system? technology? since the system `gives back'...?

% I couldn't find the missing word
Meta-signaling helps point at why exaptation through analogy-making is so powerful. We humans build mechanisms that do a lot of heavy lifting: tool-making technologies, and yes, reputation systems to enforce cooperation. When we can co-opt existing systems \cite{villani-an-agent-based-model-of-exaptive-processes-2007}, we save ourselves the effort of both \textit{building} and \textit{maintaining} a separate system. But here, something extra special is happening. The innovation was not simply horizontal -- repurposing a technology for an analogous purpose -- but vertical -- repurposing a technology for the analogous purpose \textit{of maintaining itself}. The slippage not only allows the system to perform a new task, but it also enhances its performance on its old task. This vertical co-optation is extra powerful because it sets up a positive feedback loop: enhancements to the system lead to further enhancements, since the system `gives back' to itself. For instance, in the case where tools are used for tool-making, advances in tool-making technology not only lead to better tools, but better tool-making tools, which in turn could lead to \emph{even better} tools. The same goes for compilers written in the language that they compile. For this reason, vertical co-optation seems to be at the root of leaps in complexity \cite{hofstadter-godel-escher-bach-an-eternal-golden-braid-1979}.

% Mike TODO - I don't understand the phrase `simply signaling' in this paragraph. kicking the can down the road might benefit from some examples
On a technical level, our model combines signaling with an assurance that any given choice has some probability of being observed to resolve the common problem (in evolutionary models) of continuously kicking the can down the road until we stop at a set of assumptions that one might question, either for fundamental or practical reasons. This is not to say our model has no assumptions, but rather its assumptions are as `light' as possible (at least, we have strived for this) while handling certain of the complications of kicking the can down the road by letting observers observe observers, etc. To be completely candid about the origin of this recursive structure, two reviewers in an initial round of reviews pointed out that our original model (which involved meta-signaling but no possibility of observation of meta-signalers) used meta-signaling as a \emph{deus ex machina}, questioning the realism of simply signaling (notably since signaling has no direct cost). Our new model -- that is, Model 2 described in this article -- does not contain direct costs for signaling or meta-signaling, but it does have indirect (second order) costs.

\subsubsection{Error}
Error has the advantage of keeping the `environment' varied. As social beings, other people make up a large part of our environment, and as a result, our adaptations often deal with navigating social situations. But a homogeneous population creates homogeneous social situations and correspondingly brittle social adaptations that cannot handle situations outside the monoculture. In this case, when we live in a completely homogeneous environment of cooperators, absent is the selection pressure to maintain defense mechanisms against defectors. This homogeneity problem is the reason that \textit{tit-for-tat} fails to be evolutionarily stable -- \textit{tit-for-taters} look like unconditional cooperators when interacting with each other, so cooperative (but nonconforming) strategies are indistinguishable, and can invade by drift. In our model, error, by introducing heterogeneity, provides the pressure needed to maintain defense mechanisms against potential invading strategies.

\subsubsection{Norms in communication}
This model demonstrates the benefits and dangers of communicating moral information rather than simply factual information. With the image-scoring norm, the simple morality \textit{`defection is bad, cooperation is good'} had a one-to-one correspondence between factual states of the world (cooperate/defect) and moral states (good/bad). But with the more complex stern judging norm in our model, `good' could mean cooperation with a cooperator, cooperation with someone who defected against a cooperator, defection with a someone who cooperated with someone who cooperated with a defector...

% QUESTION metasignaling is only for when observers are present? Or any signal about a third party's choices?
An infinite number of states like this collapse into `good' and `bad.' And this is the power of morality in communication: it collapses infinite factual information into (sometimes) a single bit. Agents need not know the infinite chain of who cooperated with which defectors and whether they defected against cooperators; they need only know two things: (1) what their partner did (cooperate or defect) and (2) whether that was against someone with a `good' or `bad' reputation. When Henrich objected to Boyd's recursive punishment strategy (punish those who are in bad standing, where someone is in bad standing if they fail to cooperate, or if they fail to punish someone in bad standing) \cite{henrich-cultural-group-selection-coevolutionary-processes-and-large-scale-cooperation-2004, boyd-punishment-allows-the-evolution-of-cooperation-or-anything-else-in-sizable-groups-1992}, he objected to agents having to track an infinite chain of possible transgressions -- but moral communication solves this issue, by collapsing that infinite chain at the second step. 

Embedding moral norms in the communication system comes with danger. When dealing with norms and cooperation, one can ask three questions
\begin{enumerate}
    \item Does a given norm lead to high levels of cooperation?
    \item \label{meta-question} Is that norm stable against other norms?
    \item \label{meta-question-comm} Is that norm stable against other norms \textit{when embedded in a communication system}?
\end{enumerate}

Clearly the first question is of a different class that the next two. But even questions \ref{meta-question} and \ref{meta-question-comm} deal with profoundly different dynamics. In the scenario \ref{meta-question}, someone following different norms acts according to their private rule that deviates from the prevailing norm, but otherwise minds their own business. In scenario \ref{meta-question-comm}, however, a signaler subscribing to different norms `pollutes' \textit{everyone else's} reputation information with their own normative judgment, thus impacting not only their own actions but the actions of others. This can lead to domino effects that do not exist in case \ref{meta-question}, which will therefore lead to different equilibria \cite{yamamoto-a-norm-knockout-method-on-indirect-reciprocity-to-reveal-indispensable-norms-2017}.

\paragraph{Disagreements in standing}
A reviewer remarked that we assume that there are no disagreements about an agent’s tag. This is true, but only in a weak sense, as the agreement of taggers has an explanation that is endogenous to the model: that is, the state where all agents are stern discriminators is stable, and that world is a world in which all agents agree on standing, modulo error. That `modulo error' is of course an important caveat – we do allow for disagreements, in the sense that agents can err in their tagging, and thus tag differently from how a non-erring stern discriminator would tag. However, in this setting, as long as error is not too high, the stern discriminator norm is nevertheless stable. This is partially because, unlike in cases such as image scoring, the stern discriminator norm, being aligned, does not lead to error propagation\footnote{To see why, notice that errors are “preserved” in the image scoring setting, as agents who err are then defected against in punishment, and then their punishers are subsequently defected against -– leading to a propagation of defection. In the stern discriminator case, punishers are not defected against, which mitigates the defection explosion}.

Because the model has only one observer for an action, there is only one case in the model where an agent may disagree with a tag based on her own observations: that is, if an agent disagrees with her own tag. Why, in this case, should she nonetheless signal conformingly? The simple answer is that it is potentially costly not to, as signaling non-conformingly always is (and signaling conformingly is cost-free). But perhaps this explanation for why an agent would so easily accept an erroneous tag seems overly pat –- this is an issue that could be addressed in elaborations of this model.

Of course, one might ask \emph{why} we don't allow for an action or signal being witnessed by multiple agents and getting tagged differently. Our reason is simple: the single-tagger case is in some sense the `worst-case', because multiple taggers would only increase the fidelity of the average signal, by the law of large numbers.

\subsection{Relaxation of assumptions} \label{section:discussion-relaxation}
A reasonable question might be: how much of the structure of the model is necessary for the proofs to go through? Certain parts of the model, including the way observers are assigned, are quite specific. What is actually needed?
\begin{enumerate}
\item \emph{Any} agent has a nonzero, possibly variable, probability of being observed, $p$. 
\item When deciding how to signal, agents are unaware of whether they are being observed or not (though they may know their probabilities of being observed).
\item There is some expected discounting factor, $k_\text{tot} < 1$, that accounts for a round-to-round discount rate, in addition to the possibility that one's tag might not stick, allowing the agent to escape the second order cost entirely.
\item Agents err with some probability in actions, signals, and meta-signals (leading to some effective error rate $\epsilon'$
\end{enumerate}

In this more general setting, it is still true that the cost of defecting non-conformingly, $C_\text{def}$, has the lowest cost out of all ways of not conforming, because it allows the agent to save on the first-order cost $\gamma$ of cooperation. So if we can find a set of parameters $k_\text{tot}$, $p$, and $\epsilon'$ for which $C_\text{def} > 0$, then we can conclude that it is always costly to deviate. We need therefore that

\begin{align}
&C_\text{def} = (1 - p)k_\text{tot}(1 - 2\epsilon')\beta - \gamma > 0 \\
&\Longleftrightarrow k_\text{tot} > \dfrac{\gamma}{\beta(1 - 2\epsilon')(1 - p)} \\
&\Longleftrightarrow \epsilon' < \dfrac{1}{2} - \dfrac{\gamma}{2 k_\text{tot} \beta (1 - p)} \\
&\Longleftrightarrow p < 1 - \dfrac{\gamma}{k_\text{tot}\beta(1 - 2\epsilon')}
\end{align}

In other words, with $p_\text{stick}$ sufficiently large, and $\epsilon'$ and $p$ sufficiently small (but nonzero) -- and all of them possibly variable -- it is costly to act in a way that deviates from the stern discriminator strategy, even if the exact details of the model laid out above are not adhered to.

\subsection{The strength of the equilibrium}
In the direct reciprocity case, \textit{tit-for-tat} is only collectively stable; that is, it performs at least as well as any other strategy when it is dominant, a condition weaker than evolutionary stability (for more background, see \ref{appendix:background-iterated-pd}). In fact, \emph{no} pure strategy can be evolutionarily stable in direct reciprocity models \cite{boyd-no-pure-strategy-is-evolutionarily-stable-in-the-iterated-prisoner-s-dilemma-1987}. Our equilibrium, by contrast, is \emph{stronger} than evolutionary stability \cite{smith-the-logic-of-animal-conflict-1973}, because in it, stern discriminators \emph{strictly outperform} all other strategies.

Importantly, the equilibrium we found is not the only stable point in the system. Trivially, there is the symmetrical stern reverse-discriminator state that would clearly also be stable (\texttt{cd0110}). But even if there are stable states with defection, weak group selection would be enough to select for the cooperative equilibrium. Usually, group selection is under intense time pressure, as it must quickly kill off groups that tend towards defection before they propagate. In our case, groups do not tend towards defection, since the forces we laid out produce a cooperative equilibrium. Thus, group selection no longer has such time pressure, and it need only \textit{select}, rather than \textit{maintain}, the cooperative equilibrium \cite{henrich-cultural-group-selection-coevolutionary-processes-and-large-scale-cooperation-2004}.

\subsection{The benefits of social enforcement}
We have demonstrated a mechanism for the social enforcement of truthful communication, which has a number of advantages:
\begin{enumerate}
    \item Social enforcement is more flexible in the determination of costs. This flexibility \emph{not only} allows for the the truthful signal to be the least costly, but \emph{also} to be potentially cost-free. Furthermore, it allows for more complex, combinatorial communication systems (where meaning is determined as a function of a set of symbols, rather than having a straightforward symbol-to-world-state mapping \cite{lachmann-cost-and-conflict-in-animal-signals-and-human-language-2001}), because costs can be differentially determined on the level of a statement, rather than, say, associating each symbol with a cost. This kind of flexibility would be unlikely if costs were determined physiologically, for instance. \cite{lachmann-cost-and-conflict-in-animal-signals-and-human-language-2001}
    \item It has stability benefits. When costs are not socially determined, but instead determined physiologically, there is selection pressure on the signaler to evolve ways of producing (perhaps untruthful) signals in a cheaper fashion, though there are ways to ensure that even physiological signals are honest, such as in Grafen's costly signaling model \cite{grafen_1990}. But in the social enforcement case, costs are determined by recipients and, therefore, they are less likely to be destabilized by natural selection \cite{lachmann-cost-and-conflict-in-animal-signals-and-human-language-2001}. This is not to say that they are completely immune to such disturbance, however -- one could imagine that signalers might evolve ways to evade social enforcement, by making their deceitful signals harder to verify.
\end{enumerate}

There \emph{are} constraints we place on our strategy set that make it easier for the social enforcement mechanism: for instance, signalers cannot adjust their signaling strategy based on whether they are being observed. This is not such a great problem. While there is sometimes an absence of pressure to signal in a conforming fashion, there is never any positive pressure for an individual to signal in a non-conforming fashion -- there is no benefit to smearing or praising untruthfully, because it is not the reputation of others, but ones own reputation, that determines one's fitness. Thus, even if an agent \textit{could} signal differentially based on whether they were being observed, if they even entertained any minute probability of being caught, they would always signal in a conforming fashion.

\subsection{The ubiquity of reputation}
Reputation -- and the communication system that underlies it -- is vital for all kinds of decision-making. In indirect reciprocity, it indicates who to cooperate with. In third party punishment, reputation would likely be necessary for the third party to know who to punish (see \ref{appendix:punishment} and \cite{boyd-coordinated-punishment-of-defectors-sustains-cooperation-and-can-proliferate-when-rare-2010}). And for cooperation to function as a costly signal for mate quality or alliance potential \cite{gintis-costly-signaling-and-cooperation-2001}, there must be some kind of reputation mechanism that propagates information about who is cooperating -- unless every cooperative act is observed directly by everyone, which seems unlikely.

For these cases, our results provide two alternative explanations. Either (1) a communication system for disseminating reputation information might have evolved for the indirect reciprocity case, and was then co-opted for use by third-party punishers, or (2) selection pressures analogous to the indirect reciprocity case exist in these other situations, and our research can inform their exploration. Third party punishment is especially similar to indirect reciprocity, since the latter is punishment by withholding of cooperation while the former is punishment by direct imposition of cost. While this creates a slightly different payoff matrix, it is likely that the three principles underlying the viability of communication in our case could be extended to third-party punishment. 

\subsection{Conclusions}
Human relationships cannot be reduced to pairwise interactions. Instead, we are submerged in a tangle of observation and judgment, where people form opinions and base decisions on interactions they are not involved with, using gossip to gather information we have not observed directly. As institutions, legal systems, and governments develop, the role of the third party, and therefore, communication, only increases \cite{nowak-evolution-of-indirect-reciprocity-2005, alexander-ostracism-and-indirect-reciprocity-the-reproductive-significance-of-humor-1986}. 

It is therefore vital to understand how a communication system maintaining all of this social information could possibly be stable. Here we found three sufficient conditions of stability for a simple binary-valued communication system: meta-signaling, error, and stern judging norms.

There are a number of directions to pursue in further work.
\begin{enumerate}
    \item \textit{Alternative equilibria} - We have found an equilibrium point. However, knowing the full set of equilibria, in addition to the probabilities of fixating on each given some initial conditions, will shed light on how likely a society is to attain the equilibrium we've identified. 
    \item \textit{Elaborate communication} - human language involves combining symbols to create meaning. This type of combinatorial communication lacks some of the stability properties of more basic symbol to meaning systems (like ours) and therefore needs to be analyzed further \cite{lachmann-the-disadvantage-of-combinatorial-communication-2004}. Relatedly, if communication is too elaborate, cognitive and informational bounds may place limits on communication and decision making \cite{price_jones2020}. One intriguing way to address this is to use reinforcement learning to set agent's strategies \cite{koster_etal2022}.
\end{enumerate}

\begin{center}
    \textsc{Acknowledgments}
\end{center}
We would like to thank Hajime Shimao, John Miller, and Melanie Mitchell for helping to originate many of the ideas present in this paper; Shimon Edelman, Michael Macy, and Alex Vladimirsky for their thoughts, ideas, and feedback on drafts at various stages of the writing process; and Helena Miton for both of those things. We would further like to thank our anonymous reviewers for their constructive feedback.

\begin{center}
    \textsc{Funding and Disclosure}
\end{center}
We would like to thank the NSF for funding Victor Odouard (grant number 1757923) to complete a substantial portion of this research. The final stages of this work were also supported by a grant to SFI from the Omidyar Network to cover core research in the area of Emergent Political Economy. David Krakauer is
the PI on this award. The funders played no role in the design, execution, writing, and submission of this research. We have no conflicts of interest to disclose.

\newpage

%TC:ignore
\appendix

\gdef\thesection{Appendix \Alph{section}}
\section {Background}
\subsection{Iterated Prisoner's Dilemmas} \label{appendix:background-iterated-pd}
The dominant strategy for both agents in a one-shot prisoner’s dilemma is defection --- (\texttt{d}, \texttt{d}). But if you're likely to meet your partner again, the thinking goes, it might pay to cooperate. This leads to the `iterated prisoner's dilemma.'

In defining the iterated prisoner's dilemmas, there is the further requirement that mutually cooperating should lead to a higher payoff than two partners alternating between cooperation and defection (first round, $A$ cooperates and $B$ defects, second round, the reverse, etc.). This requirement is satisfied in our version of the prisoner's dilemma, because $2(\beta - \gamma) > \beta - \gamma$.

Even here, however, defecting is an equilibrium when there is a common-knowledge upper bound on the number of iterations (shown by a backwards induction) \cite{kuhn-prisoner-s-dilemma-2019}. However, this equilibrium requires strong assumptions about human rationality \cite{bicchieri-self-refuting-theories-of-strategic-interaction-a-paradox-of-common-knowledge-1989} that, when relaxed, might allow for rational cooperation (for instance, if one player believes that the other player might act irrationally \cite{kreps-rational-cooperation-in-the-finitely-repeated-prisoner-s-dilemma-1982}).

% DONE: I'm not quite sure what "at some point" means in the following paragraph. Reword and/or discuss?
% RESOLVED did my edit clarify? 
% Yes, perfect!
One not-quite-rational strategy that does particularly well is \textit{tit-for-tat}, a short-memory reciprocity: cooperate on the first round, and cooperate with those who cooperated in the previous round. This strategy performed well in tournaments \cite{axelrod-the-emergence-of-cooperation-among-egoists-1981}, and in a world dominated by \textit{tit-for-taters}, no strategy can do better \cite{axelrod-the-evolution-of-cooperation-1981}. However, \textit{tit-for-tat} does not satisfy Maynard Smith's conditions for stability \cite{smith-the-logic-of-animal-conflict-1973}, meaning that \textit{tit-for-taters} cannot resist the population growth of alternative strategies (see Section~\ref{section:model-stability}). This is because one can always construct a competing strategy that is indistinguishable from \textit{tit-for-tat} when playing against a \textit{tit-for-tater}, and such strategies perform equally well and thus can grow in population (e.g. unconditional cooperators and \textit{tit-for-taters} are indistinguishable playing against each other, since they both cooperate all the time). In fact, a similar construction exists for any pure strategy, so no pure strategy is stable in the iterated prisoner's dilemma \cite{boyd-no-pure-strategy-is-evolutionarily-stable-in-the-iterated-prisoner-s-dilemma-1987}.

\begin{center}
    \begin{figure}[htp]
    \centering
    \includegraphics[width=9cm]
    {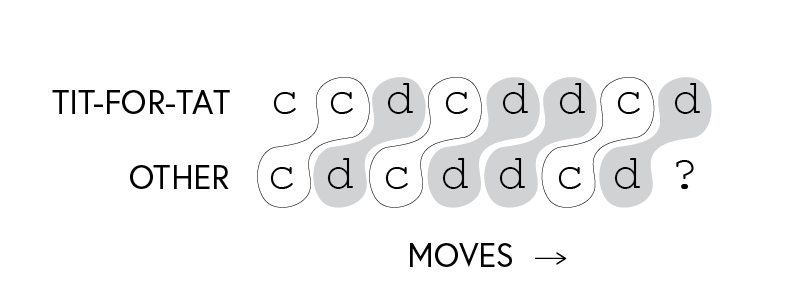}
    \caption{\textbf{Tit for taters always cooperate as many or more times as their partners.} A possible sequence of moves between a tit for tater and a randomly chosen strategy. Every \textit{tit-for-tat} defection is the result of the opponent's defection in the previous round -- so the opponent will always have defected at least as many times as the \textit{tit-for-tater}. }
    \label{fig:tit-for-tat}
\end{figure}
\end{center}

One reason \textit{tit-for-tat} performs \textit{so well} in Axelrod's tournaments, but is not evolutionarily stable, is that it can initiate long chains of mutual cooperation, where both parties do very well but where \textit{tit-for-tat} does not strictly outperform the other strategy (though enforcing a certain degree of assortativity, or relatedness between agents, can create stable cooperative states \cite{veelen-direct-reciprocity-in-structured-populations-2012}. In fact, in our prisoner's dilemma formulation, \textit{tit-for-tat} never beats another strategy head-to-head since tit-for-taters always cooperate at least as many times as their partners (Figure \ref{fig:tit-for-tat}).
% QUESTION our model doesn't support an actual tit-for-tat strategy since one cannot know if one has encountered an individual before (that being said, the probability of doing so is infinitesimal). There is a mapping we can establish if we let the experience of others substitute for our own in the signaling system. I think the nuance here was a cause for confusion in our reviewers.

\subsection{Punishment} \label{appendix:punishment}
Punishment has extensively been studied as a potential stabilizer of cooperation, partially because it is so prevalent cross-culturally \cite{henrich-costly-punishment-across-human-societies-2006, henrich-in-search-of-homo-economicus-2001}. 

Much of the literature has focused on how it can survive as a practice despite being costly to the punisher \cite{boyd-coordinated-punishment-of-defectors-sustains-cooperation-and-can-proliferate-when-rare-2010, boyd-the-evolution-of-altruistic-punishment-2003, henrich-why-people-punish-defectors-2001, boyd-punishment-allows-the-evolution-of-cooperation-or-anything-else-in-sizable-groups-1992}, though we do believe the hyper-focus on costly punishment may stem from western-centric conceptions on what punishment looks like (e.g. police spending their time and risking their lives to enforce the laws) rather than what punishment must inherently be  \cite{henrich-the-origins-and-psychology-of-cooperation-2021}. 

Still the literature is informative for the cases in which punishment is actually costly. In these cases, the problem \textit{appears} to be similar to the initial problem of costly cooperation. It is not quite the same, however, because when punishment is common, it becomes less costly, as there are fewer defectors to punish. Such is not the case with cooperation, which remains equally costly no matter how common it is \cite{boyd-the-evolution-of-altruistic-punishment-2003}. 

Partially due to this asymmetry, many solutions to the costly punishment problem have been proposed. If the punishers can recoup the costs of punishment by coercing those around them to cooperate, they can proliferate. If a recursive strategy exists that punishes non-cooperators, and punishes those that fail to punish non-cooperators, and so on, ad infinitum, then punishment can stabilize almost any behavior \cite{boyd-punishment-allows-the-evolution-of-cooperation-or-anything-else-in-sizable-groups-1992}. This is reminiscent of the behavior in our If some modicum of conformist transmission exists, that is, a tendency to absorb norms from the majority behavior of the group \cite{henrich-the-evolution-of-conformist-transmission-and-the-emergence-of-between-group-difference-1998}, punishment can also emerge \cite{henrich-why-people-punish-defectors-2001}. Furthermore, if the costs of punishment are spread across the group, as is the case for ostracism, punishment can be stable \cite{hirshleifer-cooperation-in-a-repeated-prisoners-dilemma-with-ostracism-1989}.  
% TODO: we might consider referencing our observers observing observers (etc.) here (I really mean might).

\section{Justification of approximations} \label{appendix-approximation-justifications}
To enable the derivation of mathematical results about our model, we assume an (1) infinite and (2) well-mixed population that interacts in (3) infinitely many rounds per generation. These assumptions also (counter-intuitively) have the side effect of positioning our model within the explanatory voids in previous models:
% TODO: fix section reference in first bullet
\begin{itemize}
    \item \textit{infinite rounds} - humans face cooperation choices daily, and live for tens of thousands of days. By contrast, many simulations of the evolution of cooperation simulate tens or at most hundreds of interactions per generation (e.g. \cite{smead-indirect-reciprocity-and-the-evolution-of-moral-signals-2010}, \cite{yamamoto-a-norm-knockout-method-on-indirect-reciprocity-to-reveal-indispensable-norms-2017}). For cases in which it takes time to reach a long-run equilibrium (e.g. Section~\ref{section:unexpressed-information-error}), taking the first ten or hundred rounds would not give a result representative of the long-run average. Approximating infinite rounds takes this long-run equilibrium into account much more faithfully. The assumption, of course, is not perfect -- it is true that some high-stakes cooperation decisions happen only infrequently.
    \item \textit{infinite population} - kin altruism, reciprocity, and group selection work best when group sizes are small \cite{smith-group-selection-1976}, \cite{brown-the-evolution-of-social-behavior-by-reciprocation-1982}. Because the mathematical analysis of our model is feasible for very small and very large (infinite) groups, we chose the latter in order to fill out the explanatory void for cooperation in large groups.
    \item \textit{well-mixed groups} - This means that any two individuals are equally likely to interact -- agents do not preferentially interact with a subset of the group. Many times, selective assortativity, in which cooperators preferentially interact with other cooperators, is used as an explanatory mechanism for cooperation in groups, since it reduces the payoff disadvantage of cooperators when compared to defectors. Similarly, spatial structure, in which agents interact with nearby agents, has the effect of lowering effective group size \cite{bowles-the-coevolution-of-individual-behaviors-and-social-institutions-2003, wang-cooperation-and-assortativity-with-dynamic-partner-updating-2012}. These kinds of mechanisms are generally a tailwind in favor of cooperation, and by keeping our model well-mixed, we can examine the effect of indirect reciprocity in isolation, without the help of these other forces.
\end{itemize}

In many ways, therefore, our approximations, while making the math easier, make the evolution of cooperation \textit{harder}. This is not true in all respects, however: a zero probability of invasion in the infinite case translates to a very small probability of invasion in the finite case (but this becomes negligible as $n$ becomes large).

\section{Proofs} \label{appendix-proofs}
\subsection{Cooperation can be made arbitrarily close to 100\%} \label{appendix-model2-cooperative}
Let us select a random agent, agent $A$, in an action round with her interaction partner, agent $B$. Will $A$ cooperate? There are two cases: either $B$'s tag is based on his signal (call the probability of this occurring $p_\text{s}$), in which case $B$ followed his prescribed strategy with probability $1 - \delta$, or $B$'s tag is based on his action (with probability $p_\text{a}$), in which case he followed his prescribed strategy with probability $1 - \epsilon$. Now, if $B$ followed his prescribed strategy, he will have a \texttt{1}-tag. 

If $A$ doesn't err, she will cooperate if $B$ has a \texttt{1}-tag, and if she does err, she will cooperate if $B$ has a \texttt{0}-tag. We obtain therefore that if $B$'s tag is based on his action, $A$ will cooperate with probability $(1 - \delta)(1 - \epsilon) + \delta\epsilon$, and if $B$'s tag is based on his action, $A$ will cooperate with probability $(1 - \epsilon)^2 + \epsilon^2$. $A$'s total probability of cooperating is

\begin{equation}
p_s[1 - (\epsilon + \delta - 2\epsilon\delta)] + p_a[1 - (2\epsilon - 2\epsilon^2)]
\end{equation}

Note that this approaches probability $1$ as $\epsilon$ and $\delta$ approach 0, since the two bracketed expressions approach $1$ in this limit, and $p_s + p_a = 1$ (one's tag \emph{must} either be based on one's action or one's signal). We conclude that the percentage of cooperation in this state can be made arbitrarily close to $1$.

\newpage
\bibliographystyle{apalike}
\bibliography{bibliography}
%TC:endignore
\end{document}